\documentclass{article}

\begin{document}

In Newton mechanics it is accepted that time $t$ is the same 
for
all inertial coordinate systems, i.e. it is an absolute one.
Contrary to this assumption, the Einstein's principle of
relativity is based on the fundamental idea that each inertial
coordinate system $(XYZ)_{k}$ corresponding to the given 
particle
or body is characterized with its own time $t_{k}$ . Other
important consequences from the principle of relativity, such as
a constancy of the velocity $C$ of the light, dependencies of
$M_{k}$ and $t_{k}$ values from the velocity $V_{k}$ etc. are
discussed in almost all courses of physics ( e.g.[1]).

On the other hand, one can observe-at least in the framework 
of our
Universe-that almost all particles and nuclei loose their masses 
trough the
radioactive decay. Due to different astrophysical processes 
most of
astronomical objects-such as stars, planets etc- also change 
their masses
during their ''life''. Based on these widely spread natural 
phenomena one
can formulate the following problem: would it be correct to 
assume that the
flow of time $dt$ depends from the change of mass 
$\frac{dM}{M}$ ? Can one
establish a relationship similar to the law of radioactive decay 
and apply
it to all particles and bodies? Does it mean that the following 
relationship
\begin{equation}
\frac{dM}{M}=- {\mu}dt
\end{equation}

satisfies the principle of relativity and now each particle, or
nucleus, or body is characterized with its own time? The 
answer
is positive, because time $t$ is flowing until the mass $M$
exists and $\frac{\Delta M}{M}\neq 0$. However, if $\frac{\Delta
M}{M}\rightarrow 0$, the concept of t becomes meaningless.
Unfortunately, one cannot say anything about the unknown
parameter $\mu$. It is not clear whether $\mu$ is a function of
coordinates ($XYZ$), velocity $V$, mass $M$, electrical 
charge$
Z$, etc. Although one does not know the physical meaning of 
that
parameter, one can try to extract the most simple 
consequences
from eq.(1). Also, one can make some numerical evaluations of
$\mu $, which will allow to get better understanding of problems
associated with proposed connection between $M$ and $T$.

The time interval $\Delta t=t_{2}(M_{2})-t_{1}(M_{1})$ between 
two events is
defined as

\begin{equation}
\Delta t=-{\mu}^{-1}ln(\frac{M_{2}}{M_{1}}
\end{equation}

One should note that the value of $\Delta t$ changes its sign to 
the
opposite if $M_{2}>M_{2}.$ Eq.(2) has also an exponential form:
\begin{equation}
\frac{M_{2}}{M_{1}}=exp(-\mu\Delta t)
\end{equation}

If $\mu%
\Delta t->0$, then $M_{1}=M_{2}$. Having $%
\mu\Delta t<1$, one can use a simple relationship
\begin{equation}
\Delta t={\mu}^{-1}\frac{\Delta M}{M}
\end{equation}

Eq.(4) establishes that the flow of time $t$ depends from both
the relative change of mass $\frac{\Delta M}{M}$ and the 
parameter
of $\mu$.

The velocity $V$ of particles and bodies can be found from the
following equation:
\begin{equation}
V=- {\mu} .\Delta S(\frac{\Delta M}{M})^{-1}
\end{equation}

Here $\Delta S$ is a distance. It is interesting to note that if
$\frac{\Delta M}{M}\rightarrow 0$, the flow of time becomes 
infinitely slow and thus $%
V\rightarrow\infty $. However, if $\frac{\Delta
M}{M}\rightarrow1$ the velocity $V$ reaches a certain value,
which depends from the parameter $\mu$. To satisfy the 
principle
of relativity one should assume that always $V<C$. It happens 
if
$\frac{\Delta M}{M}\neq{0}$ and if $\mu < \frac{\Delta
M}{M}$C$(\Delta S)^{-1}$.

The energy $E$ of a body with mass $M$ and velocity $V$ 
always has a
positive sign:
\begin{equation}
E=(\mu^{2}/2)M(\Delta S)^{2}\left( \frac{\Delta M}{M}\right)^{-2}
\end{equation}

Let us consider several bodies with different $M_{k}$, $(\Delta
M_{k}/M_{k}),V_{k}$ and ${\mu}_{k}$,

that are situated at different distances $\Delta S_{k}$ from the 
point they
simultaneously meet each other. It can happen if
\begin{equation}
\frac{\Delta S_{1}}{V_{1}}=\frac{\Delta 
S_{2}}{V_{2}}=...=\frac{\Delta S_{k}%
}{V_{k}}
\end{equation}

These ratios can be simplified:
\begin{equation}
\frac{1}{\mu _{1}}\frac{\Delta M_{1}}{M_{1}}=\frac{1}{\mu 
_{2}}\frac{\Delta
M_{2}}{M_{2}}=...=\frac{1}{\mu _{k}}\frac{\Delta M_{k}}{M_{k}}
\end{equation}

Thus, only moving bodies satisfying the eq.(8) can cause events
that happen simultaneously in a given point of space.

An attempt to estimate unknown parameters $\mu$ and to try 
to
understand their meaning is described below. In the case of 
decay
of nuclei and elementary particles a simple way of estimation
of$\mu_n$ is proposed. By using the eqn.(4) and well known
relationship $\Delta t\Delta E \geq(h/2\pi )$ one can reach the
following equation:
\begin{equation}{\mu}_{n}=(h/2\pi )^{-1}\Delta E\frac{\Delta M}{M}
\end{equation}

Assuming that $\Delta E\approx 1MeV$ and that $\frac{\Delta 
M}{M}%
\approx 1$ one gets an estimation of ${\mu} _{n}\approx
10^{23}sec^{-1}$. The value of $V=C$ is reached at the distance
$\Delta S$ of about

$3.10^{-13}cm$. Estimated values of $
{\mu}%
_{n}$ and $\Delta S$ are in agreement with accepted values for 
a
nuclear time of $10^{-23}sec$ and a nuclear length of
$10^{-13}cm$. \

A $\beta $-decay of neutrons is a process with week interaction
between the products of the decay. Assuming that $\frac{\Delta
M}{M}\approx 10^{-3}$ and $\Delta t\approx10^{3}sec$, one 
gets a
value of $ {\mu}_{w}\approx 10^{-6}sec^{-1}$. Most of $\beta
$-radioactive nuclei with masses around $A=100$ and half-lives 
of
about $10^{-2}-10^{2}sec$ are characterized with

values of $\frac{\Delta M}{M}\approx10^{-5}$ and values of 
$\Delta
t\approx 10^{-2}-10^{2}sec$. Therefore, one gets values of $
{\mu} _{w}\approx 10^{-3}-10^{-7}sec^{-1}$. There is no
contradiction between this interval of values and the value of
${\mu}_{w}\approx 10^{-6}sec^{-1}$ for $\beta $ -decay of
neutrons.

However, it is much more difficult to estimate the parameter 
$\mu$
g for astronomical objects, which interact with gravitational
fields of
other bodies. Assuming that a hypothetical body has a mass 
$M\;$of$%
\approx 6.10^{27}g$, a half-life $\Delta t$ of $\approx$ $%
3.(10^{16}-10^{17})sec$, that is close to the age of Universe 
and $\frac{%
\Delta M}{M}\approx $0,3.(10$^{-16}$-10$^{-17}$), one can get a
rough estimation for ${\mu} _{g}$, which is of $\approx
(10^{-33}-10^{-35})sec^{-1}.$ It is curious the normalized to
$\mu _{n}$ values of ${\mu} _{n}, {\mu} _{w}$ and ${\mu}_{g}$ to
be compared. One gets the following ratios: $
1:(10^{-26}-10^{-30}):(10^{-56}-10^{-58})$. These ratios have to
be considered only as an illustration, since we ignored large
$\beta $-decay half-lives of some nuclei that reach the value up
to $10^{15}$years. Besides that, one did not consider half-lives
of astronomical bodies that are much shorter than the age of
Universe.

Although, one does not know the physical meaning of these
parameters, it is interesting to compare them with 
dimensionless
constants $C_{S}$ , $C_{w}$ and $C_{G}$ for strong, weak and
gravitation interaction and their ratio: 1: 10$^{-24}$ :
10$^{-46}$ . Both ${\mu} $- and $C$-values are shown on Fig.1 
in
logarithmic scale and look quite similar.

As one should have expected, the proposed simple connection 
between $M$ and $%
t$ generates more problems than one could have discussed in 
this short
letter. However, an attempt to propose a new and a quite 
natural ''$M-t$''
connection, which is described above seems to be not a 
meaningless one due
to some important consequences.

References

1. L.D.Landau and E.M.Lifshitz ''Course of Theoretical 
Physics'', ''Nauka'',
Moscow, 1968.

\bigskip

\bigskip

Caption to the figure 1.

\bigskip

Comparision between the $\mu $ and $C$-values for strong, 
weak and
gravitational interactions.

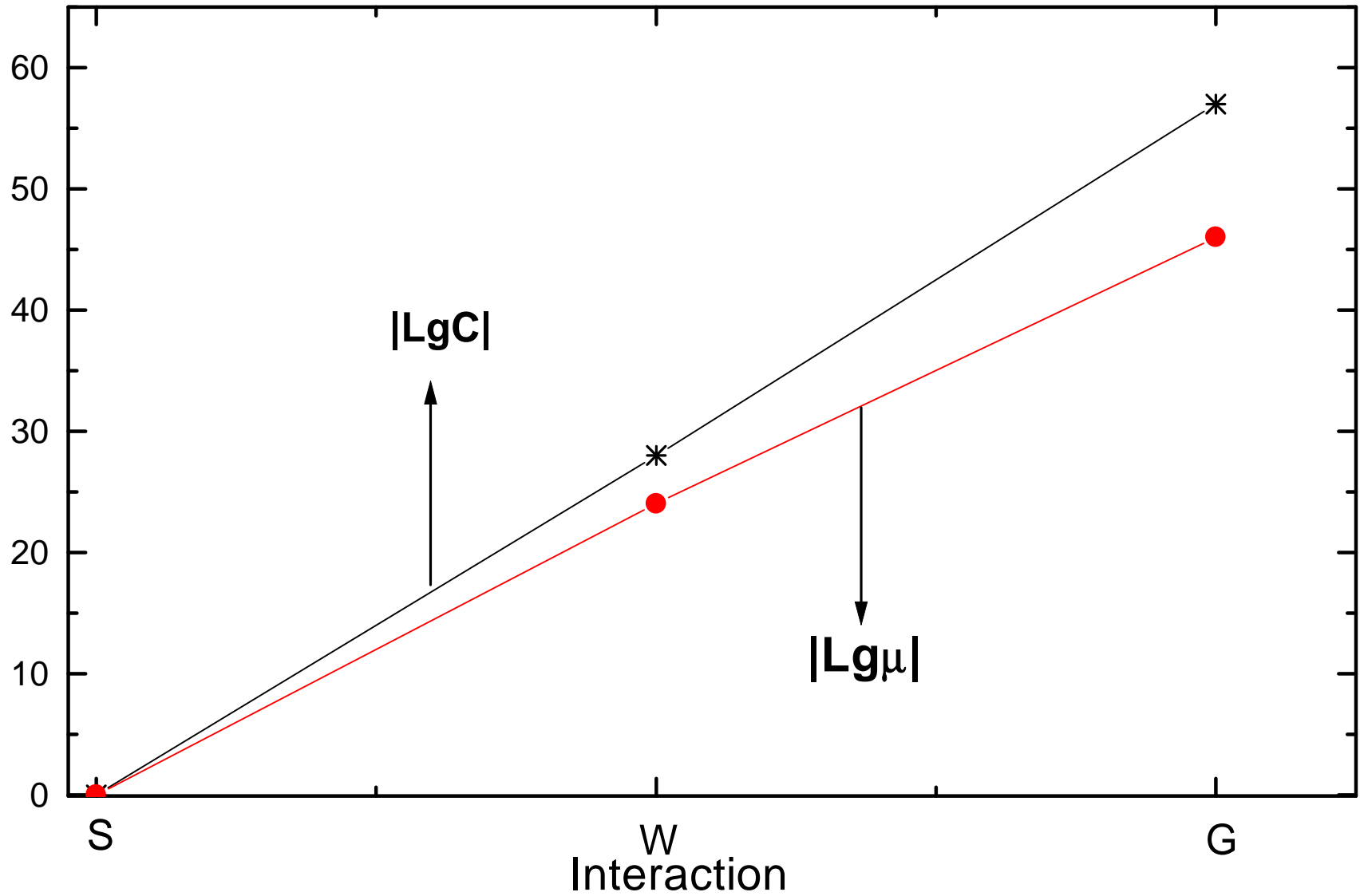
\end{document}